\documentclass [11pt,a4paper] {article}

\usepackage[cp1252]{inputenc}

\usepackage{amssymb}
\usepackage{amsmath}
\usepackage{amsfonts,amssymb}
\usepackage[dvips]{graphicx}
\usepackage{bbm}

\DeclareMathAlphabet{\mathpzc}{OT1}{pzc}{m}{it}

\newcommand*\rfrac[2]{{}^{#1}\!\!/\!_{#2}}

\begin{document}

\title{The relation between the Kochen-Specker theorem and bivalence}

\author{Arkady Bolotin\footnote{$Email: arkadyv@bgu.ac.il$\vspace{5pt}} \\ \textit{Ben-Gurion University of the Negev, Beersheba (Israel)}}

\maketitle

\begin{abstract}\noindent In the paper it is argued that the Kochen-Specker theorem necessitates a conclusion that for a quantum system it is possible to find a set of projection operators which is not truth-value bivalent; that is, a bivalent truth-value assignment function imposed on such a set cannot be total. This means that at least one proposition associated with the said set must be neither true nor false.\\

\end{abstract}

\section{Introduction}  

\noindent The main implication of the result of the Kochen-Specker (KS) theorem \cite{Bell,Kochen,Abbott} is that quantum theory fails to allow a non-contextual hidden variable model. But what is more, this theorem proves that \textit{quantum theory fails to allow a logic that obeys the principle of bivalence}.\\

\noindent To state this in more detail, let us recall that each projection operator leaves invariant any vector lying in its range, $\mathrm{ran}(\cdot)$, and annihilates any vector lying in its null space, $\mathrm{ker}(\cdot)$. Next, suppose that a quantum system is prepared in a pure state $|\Psi\rangle$ that lies in the range of the projection operator $\hat{P}_A$ associated with the proposition $A$, whose valuation is denoted by $ {[\![ A ]\!]}_v$. Being in the state $|\Psi\rangle$ is subject to the assumption that the truth-value assignment function $v$ must assign the truth value 1 (denoting \textit{the truth}) to the proposition $A$ and, thus, the operator $\hat{P}_A$, namely, $v(\hat{P}_A) = {[\![ A ]\!]}_v = 1$, since $\hat{P}_A |\Psi\rangle = 1 \cdot |\Psi\rangle$. In an analogous manner, if the system is prepared in a pure state $|\Psi\rangle$ lying in the null space of the projection operator $\hat{P}_A$, then the function $v$ must assign the truth value 0 (denoting \textit{the falsity}) to $\hat{P}_A$, namely, $v(\hat{P}_A) = {[\![ A ]\!]}_v = 0$, since $\hat{P}_A |\Psi\rangle = 0 \cdot |\Psi\rangle$.\\

\noindent Because  $\mathrm{ran}(\hat{P}_A) \neq \mathrm{ker}(\hat{P}_A)$, for any nontrivial vector $|\Psi\rangle$ it must be that $|\Psi\rangle \notin \mathrm{ker}(\hat{P}_A)$ if $|\Psi\rangle \in \mathrm{ran}(\hat{P}_A)$ as well as $|\Psi\rangle \notin \mathrm{ran}(\hat{P}_A)$ if $|\Psi\rangle \in \mathrm{ker}(\hat{P}_A)$. Thus, the definiteness of the proposition $A$ in the prepared pure state $|\Psi\rangle$ can be written down as its bivaluation, explicitly, ${[\![ A ]\!]}_v \neq 0$ if ${[\![ A ]\!]}_v = 1$ and ${[\![ A ]\!]}_v \neq 1$ if ${[\![ A ]\!]}_v = 0$.
$\,$\footnote{\label{f1}Due to the fundamental tenet of quantum theory, to represents a realizable pure state, the vector $|\Psi\rangle$ must be different from the null vector \cite{Peres}.\vspace{5pt}}
\\

\noindent Let us assume that the valuational axiom\smallskip

\begin{equation} \label{1} 
   v(\hat{P}_{\diamond})
   =
   {[\![ \diamond ]\!]}_v
   \;\;\;\;  ,
\end{equation}
\smallskip

\noindent where the symbol $\diamond$ can be replaced by any proposition (compound or simple), holds as a general principle.\\

\noindent Such an assumption brings the following question: Can the bivalent truth-values be assigned to all propositions related to the quantum system?\\

\noindent To be more specific, suppose that the sum\smallskip

\begin{equation} \label{2} 
   \sum_{i=1}^{n}{\hat{P}_{{\diamond}_i}}
   =
   \hat{1}
   \;\;\;\;  ,
\end{equation}
\smallskip

\noindent where $\hat{1}$ stands for the operator of the identity, is the resolution of the identity associated with the projection operators $\hat{P}_{{\diamond}_i}$. Then the question is, does there exist a truth-value assignment function $v$ such that if $v(\hat{1}) = 1$ then exactly one of $\hat{P}_{{\diamond}_i}$ has the truth value 1? In other words, is it possible to find a bivaluation\smallskip

\begin{equation} \label{3} 
   v(\hat{P}_{{\diamond}_i})
   =
   {[\![ {\diamond}_i ]\!]}_v
   \in
   \{ 1,0 \}
   \;\;\;\;   
\end{equation}
\smallskip

\noindent such that the following entailment\smallskip

\begin{equation} \label{4} 
   v
   \left(
   \sum_{i=1}^{n}{\hat{P}_{{\diamond}_i}}
   \right)
   =
   1
   \quad
   \implies
   \quad
   \sum_{i=1}^{n}{v(\hat{P}_{{\diamond}_i})}
   =
   1
   \;\;\;\;   
\end{equation}
\smallskip

\noindent hold even before the measurement of $\hat{P}_{{\diamond}_i}$ (i.e., the verification of ${\diamond}_i$)?\\

\noindent The KS theorem shows that the answer is no for a system whose Hilbert space $\mathcal{H}$ has dimension greater than two. This answer can be interpreted as showing that prior to their verification the propositions related to the quantum system do not obey a bivalent logic, which means that at least one of them must be \textit{neither true nor false}.\\

\noindent Let us present such an interpretation of the KS theorem in this paper.\\

\section{Preliminaries}  

\noindent Consider the Hilbert space $\mathcal{H} = \mathbb{C}^{4 \times 4}$ formed by complex $4 \times 4$ matrices related to the states for the spin of the composite system containing two spin--$\rfrac{1}{2}$ particles, namely,\smallskip

\begin{equation} \label{5} 
   |{\Psi}_{j\alpha k\beta}\rangle
   =
   |{\Psi}_{j\alpha}^{(1)}\rangle
   \otimes
   |{\Psi}_{k\beta}^{(2)}\rangle
   \;\;\;\;  ,
\end{equation}
\smallskip

\noindent where $j$ and $k$ are elements of the set $\{x,y,z\}$, $\alpha$ and $\beta$ are elements of the set $\{+,-\}$, $|{\Psi}_{\cdot\cdot}^{(\cdot)}\rangle$ represent the normalized eigenvectors of the Pauli matrices for each particle.\\

\noindent Let $\mathcal{O}$ be a set of 12 projection operators $\hat{P}_{j\alpha\beta}$ on $\mathbb{C}^{4 \times 4}$ which are defined by\smallskip

\begin{equation} \label{6} 
   \hat{P}_{j\alpha\beta}
   =
   |{\Psi}_{j\alpha}^{(1)}\rangle \langle {\Psi}_{j\beta}^{(1)}|
   \otimes
   |{\Psi}_{j\alpha}^{(2)}\rangle \langle {\Psi}_{j\beta}^{(2)}|
   \;\;\;\;   
\end{equation}
\smallskip

\noindent and associated with the propositions $J_{\alpha\beta}$ in a way that\smallskip

\begin{equation} \label{7} 
   v(\hat{P}_{j\alpha\beta})
   =
   {[\![ J_{\alpha\beta} ]\!]}_v
   \;\;\;\;  ,
\end{equation}
\smallskip

\noindent where $J \in \{X,Y,Z\}$.\\

\noindent Let the set $\mathcal{O}$ be separated into three subsets: $\mathcal{C}_z = \{ \hat{P}_{z\alpha\beta}\}$, $\mathcal{C}_x = \{ \hat{P}_{x\alpha\beta}\}$ and $\mathcal{C}_y = \{ \hat{P}_{y\alpha\beta}\}$, each called \textit{a context}, explicitly,\smallskip 

\begin{equation} \label{8} 
   \mathcal{C}_z
   =
   \left\{
      \left[
         \begin{array}{r r r r}
            1 & 0 & 0 & 0\\
            0 & 0 & 0 & 0\\
            0 & 0 & 0 & 0\\
            0 & 0 & 0 & 0
         \end{array}
      \right]
      \!
      ,
      \!
      \left[
         \begin{array}{r r r r}
            0 & 0 & 0 & 0\\
            0 & 1 & 0 & 0\\
            0 & 0 & 0 & 0\\
            0 & 0 & 0 & 0
         \end{array}
       \right]
      \!
      ,
      \!
      \left[
         \begin{array}{r r r r}
            0 & 0 & 0 & 0\\
            0 & 0 & 0 & 0\\
            0 & 0 & 1 & 0\\
            0 & 0 & 0 & 0
         \end{array}
      \right]
      \!
      ,
      \!
      \left[
         \begin{array}{r r r r}
            0 & 0 & 0 & 0\\
            0 & 0 & 0 & 0\\
            0 & 0 & 0 & 0\\
            0 & 0 & 0 & 1
         \end{array}
      \right]
   \right\}
   \;\;\;\;  ,
\end{equation}

\begin{equation} \label{9} 
   \mathcal{C}_x
   =
   \left\{
      \frac{1}{4}\!\!
      \left[
         \begin{array}{r r r r}
            1 & 1 & 1 & 1\\
            1 & 1 & 1 & 1\\
            1 & 1 & 1 & 1\\
            1 & 1 & 1 & 1
         \end{array}
      \right]
      \!
      ,
      \!
      \frac{1}{4}\!\!
      \left[
         \begin{array}{r r r r}
             1 & -1 &  1 & -1\\
            -1 &  1 & -1 &  1\\
             1 & -1 &  1 & -1\\
            -1 &  1 & -1 &  1
         \end{array}
      \right]
      \!
      ,
      \!
      \frac{1}{4}\!\!
      \left[
         \begin{array}{r r r r}
             1 &  1 & -1 & -1\\
             1 &  1 & -1 & -1\\
            -1 & -1 &  1 &  1\\
            -1 & -1 &  1 &  1
         \end{array}
      \right]
      \!
      ,
      \!
      \frac{1}{4}\!\!
      \left[
         \begin{array}{r r r r}
             1 & -1 & -1 &  1\\
            -1 &  1 &  1 & -1\\
            -1 &  1 &  1 & -1\\
             1 & -1 & -1 &  1
         \end{array}
      \right]
   \right\}
   \;\;\;\;  ,
\end{equation}

\begin{equation} \label{10} 
   \mathcal{C}_y
   =
   \left\{
      \frac{1}{4}\!\!
      \left[
         \begin{array}{r r r r}
             1 & -i & -i & -1\\
             i &  1 &  1 & -i\\
             i &  1 &  1 & -i\\
            -1 &  i &  i &  1
         \end{array}
      \right]
      \!
      ,
      \!
      \frac{1}{4}\!\!
      \left[
         \begin{array}{r r r r}
             1 &  i & -i &  1\\
            -i &  1 & -1 & -i\\
             i & -1 &  1 &  i\\
             1 &  i & -i &  1
         \end{array}
      \right]
      \!
      ,
      \!
      \frac{1}{4}\!\!
      \left[
         \begin{array}{r r r r}
             1 & -i &  i &  1\\
             i &  1 & -1 &  i\\
            -i & -1 &  1 & -i\\
             1 & -i &  i &  1
         \end{array}
      \right]
      \!
      ,
      \!
      \frac{1}{4}\!\!
      \left[
         \begin{array}{r r r r}
             1 &  i &  i & -1\\
            -i &  1 &  1 &  i\\
            -i &  1 &  1 &  i\\
            -1 & -i & -i &  1
         \end{array}
      \right]
   \right\}
   \;\;\;\;  .
\end{equation}
\smallskip

\noindent It is not difficult to see that within each context $\mathcal{C}_j$ the projection operators $\hat{P}_{j\alpha\beta}$ give the resolution of the identity\smallskip

\begin{equation} \label{11} 
   \sum_{\alpha\beta}{\hat{P}_{j\alpha\beta}}
   =
   {\hat{1}}_4
   \;\;\;\;  ,
\end{equation}
\smallskip

\noindent and, additionally, their product is a projection operator as well, namely,\smallskip

\begin{equation} \label{12} 
   \hat{P}_{j\alpha\beta} \cdot \hat{P}_{j\gamma\delta}
   =
   \hat{P}_{j\gamma\delta} \cdot \hat{P}_{j\alpha\beta}
   =
   {\hat{0}}_4
   \;\;\;\;  ,
\end{equation}
\smallskip

\noindent where ${\hat{1}}_4$ and ${\hat{0}}_4$ are the identity and zero matrices, respectively, $\gamma$ and $\delta$ are elements of the set $\{+,-\}$ different from $\alpha$ and $\beta$, respectively.\\

\noindent In contrast, the projection operators $\hat{P}_{j\alpha\beta}$ and $\hat{P}_{k\epsilon\zeta}$ taken from different contexts $\mathcal{C}_j$ and $\mathcal{C}_k$, where $\epsilon$ and $\zeta$ are elements of the set $\{+,-\}$ (not necessarily different from $\alpha$ and $\beta$), do not commute, that is,\smallskip

\begin{equation} \label{13} 
   \hat{P}_{j\alpha\beta} \cdot \hat{P}_{k\epsilon\zeta}
   \neq
   \hat{P}_{k\epsilon\zeta} \cdot \hat{P}_{j\alpha\beta}
   \;\;\;\;  ,
\end{equation}
\smallskip

\noindent therefore, neither $\hat{P}_{j\alpha\beta} \cdot \hat{P}_{k\epsilon\zeta}$ nor $\hat{P}_{k\epsilon\zeta} \cdot \hat{P}_{j\alpha\beta}$ is a projection operator on $\mathbb{C}^{4 \times 4}$.\\

\noindent Let us introduce a lattice $L(\mathbb{C}^{4 \times 4})$ of the subspaces of $\mathbb{C}^{4 \times 4}$, specifically, $\mathrm{ran}(\hat{P}_{\diamond})$, where the partial order $\le$ is the inclusion relation $\subseteq$ on a set of $\mathrm{ran}(\hat{P}_{\diamond})$, the meet $\sqcap$ is their intersection $\cap$ and the join $\sqcup$ is the span of their union $\cup$. The lattice $L(\mathbb{C}^{4 \times 4})$ is bounded, with the trivial space $\{0\}$ equal to the range of the zero matrix, $\mathrm{ran}({\hat{0}}_4)=\{0\}$, as the bottom and the space $\mathbb{C}^{4 \times 4}$ equal to the range of the identity matrix, $\mathrm{ran}({\hat{1}}_4)=\mathbb{C}^{4 \times 4}$, as the top.\\

\noindent Given that there is a one-to-one correspondence between $\mathrm{ran}(\hat{P}_{\diamond})$ and the corresponding projection operators $\hat{P}_{\diamond}$, one can take $\hat{P}_{\diamond}$ to be the elements of $L(\mathbb{C}^{4 \times 4})$.\\

\noindent Specifically, as\smallskip

\begin{equation} \label{14} 
   \mathrm{ran}(\hat{P}_{j\alpha\beta})
   \subseteq
   \mathrm{ker}(\hat{P}_{j\gamma\delta})
   =
   \mathrm{ran}({\hat{1}}_4 - \hat{P}_{j\gamma\delta})
   \;\;\;\;  ,
\end{equation}
\smallskip

\noindent one can define the partial order $\le$ within each context $\mathcal{C}_j \subset \mathcal{O}$ by setting\smallskip

\begin{equation} \label{15} 
   \hat{P}_{j\alpha\beta}
   \le
   ({\hat{1}}_4 - \hat{P}_{j\gamma\delta})
   \quad
   \text{iff}
   \quad
   \hat{P}_{j\alpha\beta}
   \sqcap
   ({\hat{1}}_4 - \hat{P}_{j\gamma\delta})
   =
   \hat{P}_{j\alpha\beta}
   \;\;\;\;  ,
\end{equation}
\smallskip

\noindent which means that the meet of $\hat{P}_{j\alpha\beta}$ and $\hat{P}_{j\gamma\delta}$ in $L(\mathbb{C}^{4 \times 4})$ can be defined as\smallskip

\begin{equation} \label{16} 
   \hat{P}_{j\alpha\beta}
   \sqcap
   \hat{P}_{j\gamma\delta}
   =
   \hat{P}_{j\alpha\beta} \cdot \hat{P}_{j\gamma\delta}
   =
   {\hat{0}}_4
   \;\;\;\;  .
\end{equation}
\smallskip

\noindent Since the subspaces $\mathrm{ran}(\hat{P}_{j\alpha\beta})$ and $\mathrm{ran}(\hat{P}_{j\gamma\delta})$ satisfy the following property\smallskip

\begin{equation} \label{17} 
   \mathrm{ran}(\hat{P}_{j\alpha\beta})
   \cap
   \mathrm{ran}(\hat{P}_{j\gamma\delta})
   =
   \mathrm{ran}(\hat{P}_{j\alpha\beta} \cdot \hat{P}_{j\gamma\delta})
   =
   \mathrm{ran}({\hat{0}}_4)
   =
   \{0\}
   \;\;\;\;  ,
\end{equation}
\smallskip

\noindent the join of the projection operators taken from the same context can be defined as their sum, i.e.,\smallskip

\begin{equation} \label{18} 
   \bigsqcup_{\alpha\beta}
   \hat{P}_{j\alpha\beta}
   =
   \sum_{\alpha\beta}{\hat{P}_{j\alpha\beta}}
   =
   {\hat{1}}_4
   \;\;\;\;  .
\end{equation}
\smallskip

\noindent As the identity matrix ${\hat{1}}_4$ leaves invariant any column vector $|\Psi\rangle$ lying in the space $\mathbb{C}^{4 \times 4}$, the range of ${\hat{1}}_4$, a proposition represented by ${\hat{1}}_4$ must be true in any state of the system, i.e., such a proposition must be a tautology $\top$. Also, as the zero matrix ${\hat{0}}_4$ annihilates any column vector in $\mathbb{C}^{4 \times 4}$, the null space of ${\hat{0}}_4$, a proposition represented by ${\hat{0}}_4$ must be false in any state of the system, in other words, this proposition must be a contradiction $\bot$.\\

\noindent This can be written as\smallskip

\begin{equation} \label{19} 
   |{\Psi}\rangle
   \in
   \mathrm{ran}({\hat{1}}_4) = \mathbb{C}^{4 \times 4}
   \quad
   \implies
   \quad
   v({\hat{1}}_4)
   =
   {[\![ \top ]\!]}_v
   =
   1
   \;\;\;\;  ,
\end{equation}

\begin{equation} \label{20} 
   |{\Psi}\rangle
   \in
   \mathrm{ker}({\hat{0}}_4) = \mathbb{C}^{4 \times 4}
   \quad
   \implies
   \quad
   v({\hat{0}}_4)
   =
   {[\![ \bot ]\!]}_v
   =
   0
   \;\;\;\;  .
\end{equation}
\smallskip

\noindent In keeping with the valuational axiom (\ref{1}), let us assume that conjunction and disjunction on the propositions $J_{\alpha\beta}$ represented by the projection operators taken from the same context $\mathcal{C}_j$ are defined respectively as\smallskip

\begin{equation} \label{21} 
   v\!\left(
   \hat{P}_{j\alpha\beta}
   \sqcap
   \hat{P}_{j\gamma\delta}
   \right)
   =
   v({\hat{0}}_4)
   =
   {[\![ J_{\alpha\beta} \land J_{\gamma\delta} ]\!]}_v
   =
   0
   \;\;\;\;  ,
\end{equation}

\begin{equation} \label{22} 
   v\Big(
   \bigsqcup_{\alpha\beta}
   \hat{P}_{j\alpha\beta}
   \Big)
   =
   v({\hat{1}}_4)
   =
   \Big[\!\!\Big[ {\bigvee_{\alpha\beta} J_{\alpha\beta}} \Big]\!\!\Big]_v
   =
   1
   \;\;\;\;  .
\end{equation}
\smallskip

\section{Logical account of the KS theorem}  

\noindent Imagine that the pair of spin--$\rfrac{1}{2}$ particles is prepared in a correlated spin state $|{\Psi}_{j\alpha k\beta}\rangle$ where $j=k$, say, such one that is represented by the column vector $|\Psi_{z++}\rangle$\smallskip

\begin{equation} \label{23} 
   |{\Psi}_{z++}\rangle
   =
   |{\Psi}_{z+}^{(1)}\rangle
   \otimes
   |{\Psi}_{z+}^{(2)}\rangle
   =
   \left[
      \begin{array}{r}
         1\\
         0\\
         0\\
         0
      \end{array}
   \right]
   \;\;\;\;   
\end{equation}
\smallskip

\noindent sitting in the range of the projection operator $\hat{P}_{z++}$\smallskip

\begin{equation} \label{24} 
   \mathrm{ran}(\hat{P}_{z++})
   =
   \left\{
   \left[
      \begin{array}{r}
         a\\
         0\\
         0\\
         0
      \end{array}
   \right]
   \!:
   \;
   a \in \mathbb{R}
   \right\}
   \;\;\;\;   
\end{equation}
\smallskip

\noindent and the null spaces of the projection operators $\hat{P}_{z+-}$,$\hat{P}_{z-+}$ and $\hat{P}_{z--}$

\begin{equation} \label{25} 
   \mathrm{ker}(\hat{P}_{z+-})
   =
   \left\{
   \left[
      \begin{array}{r}
         a\\
         0\\
         b\\
         c
      \end{array}
   \right]
   \!:
   \;
   a,b,c \in \mathbb{R}
   \right\}
    \;\;\;\;  ,
\end{equation}

\begin{equation} \label{26} 
   \mathrm{ker}(\hat{P}_{z-+})
   =
   \left\{
   \left[
      \begin{array}{r}
         a\\
         b\\
         0\\
         c
      \end{array}
   \right]
   \!:
   \;
   a,b,c \in \mathbb{R}
   \right\}
   \;\;\;\;  ,
\end{equation}

\begin{equation} \label{27} 
   \mathrm{ker}(\hat{P}_{z--})
   =
   \left\{
   \left[
      \begin{array}{r}
         a\\
         b\\
         c\\
         0
      \end{array}
   \right]
   \!:
   \;
   a,b,c \in \mathbb{R}
   \right\}
   \;\;\;\;  .
\end{equation}
\smallskip

\noindent Clearly, being in the state $|\Psi_{z++}\rangle$ causes all the propositions $Z_{\alpha\beta}$ associated with the context $\mathcal{C}_z$ become bivalent, that is,\smallskip

\begin{equation} \label{28} 
   |\Psi_{z++}\rangle
   \in
   \left\{
      \begin{array}{r}
         \mathrm{ran}(\hat{P}_{z++})\\
         \mathrm{ker}(\hat{P}_{z+-})\\
         \mathrm{ker}(\hat{P}_{z-+})\\
         \mathrm{ker}(\hat{P}_{z--})
      \end{array}
   \right.
   \quad
   \implies
   \quad
   \left\{
      \begin{array}{r}
         v(\hat{P}_{z++})={[\![ Z_{++} ]\!]}_v=1\\
         v(\hat{P}_{z+-})={[\![ Z_{+-} ]\!]}_v=0\\
         v(\hat{P}_{z-+})={[\![ Z_{-+} ]\!]}_v=0\\
         v(\hat{P}_{z--})={[\![ Z_{--} ]\!]}_v=0
      \end{array}
   \right.
   \;\;\;\;  .
\end{equation}
\smallskip

\noindent One can infer from here that in any correlated spin state $|\Psi_{j\alpha\beta}\rangle$ there is a context $\mathcal{C}_j$ such that among all the propositions $J_{\alpha\beta}$ associated with $\mathcal{C}_j$ one is true while the others are false, and, hence, the entailment (\ref{4}) is valid:\smallskip

\begin{equation} \label{29} 
   |\Psi_{j\alpha\beta}\rangle
   \in
   \left\{
      \begin{array}{l}
         \mathrm{ran}(\hat{P}_{j\alpha\beta})\\
         \mathrm{ker}(\hat{P}_{j\gamma\delta})
      \end{array}
   \right.
   \quad
   \implies
   \quad
   \left\{
      \begin{array}{l}
         v(\hat{P}_{j\alpha\beta})    ={[\![ J_{\alpha\beta} ]\!]}_v=1\\
         v(\hat{P}_{j\gamma\delta})={[\![ J_{\gamma\delta} ]\!]}_v=0
      \end{array}
   \right.
   \;\;\;\;  .
\end{equation}
\smallskip

\noindent As follows, in the said context $\mathcal{C}_j$, the truth values of conjunctions and disjunctions can be expressed with the basic operations of arithmetic or by the $\min$ and $\max$ functions, namely,\smallskip

\begin{equation} \label{30} 
   {[\![ J_{\alpha\beta} \land J_{\gamma\delta} ]\!]}_v
   =
   {[\![ J_{\alpha\beta} ]\!]}_v
   \cdot
   {[\![ J_{\gamma\delta} ]\!]}_v
   =
   \min\left\{
   {[\![ J_{\alpha\beta} ]\!]}_v
   ,
   {[\![ J_{\gamma\delta} ]\!]}_v
   \right\}
   =
   0
   \;\;\;\;  ,
\end{equation}

\begin{equation} \label{31} 
   \Big[\!\!\Big[ {\bigvee_{\alpha\beta} J_{\alpha\beta}} \Big]\!\!\Big]_v
   =
   \sum_{\alpha\beta}{{[\![ J_{\alpha\beta} ]\!]}_v}   
   =
   \max_{\alpha\beta}\left\{
   {[\![ J_{\alpha\beta} ]\!]}_v
   \right\}
   =
   1
   \;\;\;\;  .
\end{equation}
\smallskip

\noindent Consider a projection operator on $\mathbb{C}^{4 \times 4}$ that is not an element of the preselected (by the preparation of the composite system's  state) context $\mathcal{C}_z$: Take, for example, the operator $\hat{P}_{y++}$ whose range and null space are as follows:\smallskip

\begin{equation} \label{32} 
   \mathrm{ran}(\hat{P}_{y++})
   =
   \left\{
   \left[
      \begin{array}{r}
           a\\
          ia\\
          ia\\
         -a
      \end{array}
   \right]
   \!:
   \;
   a \in \mathbb{C}
   \right\}
   \;\;\;\;  ,
\end{equation}

\begin{equation} \label{33} 
   \mathrm{ker}(\hat{P}_{y++})
   =
   \left\{
   \left[
      \begin{array}{r}
         ia+ib+c\\
                   a\\
                   b\\
                   c
      \end{array}
   \right]
   \!:
   \;
   a,b,c \in \mathbb{C}
   \right\}
      \;\;\;\;  .
\end{equation}
\smallskip

\noindent If ${[\![ Z_{++} ]\!]}_v = 1$, then the proposition $Y_{++}$ cannot be bivalent under the truth-value assignment function $v$, otherwise one would get a contradiction $1=0$, to be exact,\smallskip

\begin{equation} \label{34} 
   \left[
      \begin{array}{r}
         1\\
         0\\
         0\\
         0
      \end{array}
   \right]
   \in
   \left\{
   \left[
      \begin{array}{r}
           a\\
          ia\\
          ia\\
         -a
      \end{array}
   \right]
   \!:
   \;
   a \in \mathbb{C}
   \right\}
   \;\;\;\;  ,
\end{equation}

\begin{equation} \label{35} 
   \left[
      \begin{array}{r}
         1\\
         0\\
         0\\
         0
      \end{array}
   \right]
   \in
   \left\{
   \left[
      \begin{array}{r}
         ia+ib+c\\
                   a\\
                   b\\
                   c
      \end{array}
   \right]
   \!:
   \;
   a,b,c \in \mathbb{C}
   \right\}
   \;\;\;\;  .
\end{equation}
\smallskip

\noindent The same contradiction would obviously appear for any other projection operator on $\mathbb{C}^{4 \times 4}$ not belonging to $\mathcal{C}_z$. Thus, in any correlated spin state $|\Psi_{j\alpha\beta}\rangle$ there is only one context $\mathcal{C}_j$ in the set $\mathcal{O}$ where the sole projection operator can be assigned the truth-value $1$ at the same time as all the rest are assigned the truth-value $0$.\\

\noindent Next, imagine that the pair of spin--$\rfrac{1}{2}$ particles is prepared in an uncorrelated spin state $|{\Psi}_{j\alpha k\beta}\rangle$ where $j \neq k$. Due to the one-to-one correspondence between column spaces $\mathrm{ran}(\cdot)$ and projection operators, the column vector $|{\Psi}_{j\alpha k\beta}\rangle$ does not lie in the column or null space of either projection operator $\hat{P}_{j\alpha\beta}$, which implies that in this case no projection operator in the set $\mathcal{O}$ can be bivalent under $v$, specifically,\smallskip

\begin{equation} \label{36} 
   j \neq k
   :
   \;\;
   |{\Psi}_{j\alpha k\beta}\rangle
   \notin
   \left\{
      \begin{array}{r}
         \mathrm{ran}(\hat{P}_{j\alpha\beta})\\
         \mathrm{ker}(\hat{P}_{j\alpha\beta})
      \end{array}
   \right.
   \quad
   \implies
   \quad
   \left\{
      \begin{array}{r}
         v(\hat{P}_{j\alpha\beta})={[\![ J_{\alpha\beta} ]\!]}_v \neq 1\\
         v(\hat{P}_{j\alpha\beta})={[\![ J_{\alpha\beta} ]\!]}_v \neq 0
      \end{array}
   \right.
   \;\;\;\;  .
\end{equation}
\smallskip

\noindent This means that independently of the state $|{\Psi}_{j\alpha k\beta}\rangle$ in which the pair of spin--$\rfrac{1}{2}$ particles can be prepared, the set $\mathcal{O}$ cannot be truth-value bivalent under $v$; otherwise stated, $v$ cannot be a total two-valued function, namely,\smallskip

\begin{equation} \label{37} 
   v
   :
   \mathcal{O}
   \nrightarrow
   \{1,0\}
   \;\;\;\;  .
\end{equation}
\smallskip

\section{Concluding remarks}  

\noindent The fact that the truth-value assignment function $v$ imposed on the set $\mathcal{O}$ can be only partial indicates that unless they are associated with the preselected context, prior to their verification the propositions $J_{\alpha\beta}$ have either a truth-value $\mathfrak{v}$ different from 1 and 0, explicitly,\smallskip

\begin{equation} \label{38} 
   v(\hat{P}_{j\alpha\beta})
   =
   {[\![ J_{\alpha\beta} ]\!]}_v
   \in
   \{
      0 < \mathfrak{v} < 1
      \,|\,
      \mathfrak{v} \in \mathbb{R}
   \}
   \;\;\;\;  ,
\end{equation}
\smallskip

\noindent or absolutely no truth-value, that is,\smallskip

\begin{equation} \label{39} 
   \left\{
   v(\hat{P}_{j\alpha\beta})
   \right\}
   =
   \left\{
   {[\![ J_{\alpha\beta} ]\!]}_v
   \right\}
   =
   \emptyset
   \;\;\;\;  .
\end{equation}
\smallskip

\noindent In the first case, prior to the verification the propositions $J_{\alpha\beta}$ obey many-valued semantics, for example, the {\L}ukasiewicz-Pykacz model of infinite-valued logic \cite{Pykacz95,Pykacz15b}. Within this model, the entailment (\ref{4}) fails because $v(\sum_{\alpha\beta}\hat{P}_{j\alpha\beta})=1$ means that $\sum_{\alpha\beta}v(\hat{P}_{j\alpha\beta}) \ge 1$, i.e., more than one $\hat{P}_{j\alpha\beta}$ can have non-zero truth-value.\\

\noindent In contrast, in the second case, before the verification the propositions $J_{\alpha\beta}$ comply with partial semantics having truth-value gaps, such as supervaluationism \cite{Varzi,Keefe}. According to supervaluationism, the entailment (\ref{4}) fails because $\bigvee_{{\alpha\beta}}J_{\alpha\beta}$ should be true regardless of whether or not its disjuncts $J_{\alpha\beta}$ have a truth value (supervaluationism describes $\bigvee_{{\alpha\beta}}J_{\alpha\beta}$ as ``supertrue'').\\

\noindent Mathematically though, partial semantics are not very different from many-valued semantics. Moreover, for any partial (``gappy'') semantics, one can construct a gapless many-valued semantics which will define the same logic \cite{Beziau}.\\

\bibliographystyle{References}
\bibliography{References}

\end{document}